\documentclass[traditabstract,longauth]{aa}

\usepackage{graphicx}

\usepackage{txfonts}

\usepackage{natbib}
\bibpunct{(}{)}{;}{a}{}{,} % to follow the A&A style
\defcitealias{leger2009}{L{\'e}ger, Rouan, Schneider et al. 2009}

\def\ms{\hbox{\,m\,s$^{-1}$}}         %m.s-1
\def\m2s2{\hbox{\,m$^{2}$\,s$^{-2}$}} %m2.s-2
\def\kms{\hbox{\,km\,s$^{-1}$}}       %km.s-1
\def\gcm3{\hbox{\,g\,cm$^{-3}$}}      %g.cm-3
\def\vsini{\hbox{$v$\,sin\,$i$}}      %vsini
\def\Mjup{\hbox{$\mathrm{M}_{\rm Jup}$}}
\def\Rjup{\hbox{$\mathrm{R}_{\rm Jup}$}}

\def\degr{\hbox{$^\circ$}}

\begin{document}

% http://www.eso.org/sci/observing/policies/publications.html
\title{Transiting exoplanets from the CoRoT space mission\thanks{The CoRoT space mission, launched on December 27th 2006, has
  been developed and is operated by CNES, with the contribution of
  Austria, Belgium, Brazil, ESA (RSSD and Science Programme), Germany
  and Spain. 
% CFHT
  Part of the observations were obtained at the Canada-France-Hawaii
  Telescope (CFHT) which is operated by the National Research Council
  of Canada, the Institut National des Sciences de l'Univers of the
  Centre National de la Recherche Scientifique of France, and the
  University of Hawaii.
% HARPS@ESO
  Based on observations made with HARPS spectrograph on the
  3.6-m European Organisation for Astronomical Research in the
  Southern Hemisphere telescope at La Silla Observatory, Chile (ESO
  program 184.C-0639).
% IAC80
  Based on observations made with the IAC80 telescope
  operated on the island of Tenerife by the Instituto de
  Astrof{\'i}sica de Canarias in the Spanish Observatorio del Teide.
% Keck
  Part of the data presented herein were obtained at the W.M. Keck
  Observatory, which is operated as a scientific partnership among the
  California Institute of Technology, the University of California and
  the National Aeronautics and Space Administration. The Observatory
  was made possible by the generous financial support of the W.M. Keck
  Foundation. 
  }
}

\subtitle{XVII. The hot Jupiter CoRoT-17b: a very old planet} 

\author{Sz.~Csizmadia\inst{\ref{DLR}}
\and C.~Moutou\inst{\ref{LAM}} 
\and M.~Deleuil\inst{\ref{LAM}} 
\and J.~Cabrera\inst{\ref{DLR},\ref{LUTh}} 
\and M.~Fridlund\inst{\ref{ESA}}
\and D.~Gandolfi\inst{\ref{ESA}}
\and S.~Aigrain\inst{\ref{Oxford}} 
\and R.~Alonso\inst{\ref{Geneve}} 
\and J.-M.~Almenara\inst{\ref{LAM}} 
\and M.~Auvergne\inst{\ref{LESIA}} 
\and A.~Baglin\inst{\ref{LESIA}}
\and P.~Barge\inst{\ref{LAM}} 
\and A.~S.~Bonomo\inst{\ref{LAM}}  
\and P.~Bord\'e\inst{\ref{IAS}} 
\and F.~Bouchy\inst{\ref{OHP},\ref{IAP}} 
\and H.~Bruntt\inst{\ref{LESIA}}
\and L.~Carone\inst{\ref{Koeln}} 
\and S.~Carpano\inst{\ref{ESA}} 
\and C.~Cavarroc\inst{\ref{IAS}}
\and W.~Cochran\inst{\ref{McDonald}}
\and H.~J.~Deeg\inst{\ref{IAC},\ref{Laguna}} 
\and R.~F.~D{\'i}az\inst{\ref{IAP}}
\and R.~Dvorak\inst{\ref{Wien}} 
\and M.~Endl\inst{\ref{McDonald}}
\and A.~Erikson\inst{\ref{DLR}}
\and S.~Ferraz-Mello\inst{\ref{Brasil}} 
\and Th.~Fruth\inst{\ref{DLR}}
\and J.-C.~Gazzano\inst{\ref{LAM},\ref{OCA}}
\and M.~Gillon\inst{\ref{Liege}} 
\and E.~W.~Guenther\inst{\ref{Tautenburg}} 
\and T.~Guillot\inst{\ref{OCA}} 
\and A.~Hatzes\inst{\ref{Tautenburg}} 
\and M.~Havel\inst{\ref{OCA}}
\and G.~H\'ebrard\inst{\ref{OHP},\ref{IAP}} 
\and E.~Jehin\inst{\ref{Liege}}
\and L.~Jorda\inst{\ref{LAM}} 
\and A.~L\'eger\inst{\ref{IAS}} 
\and A.~Llebaria\inst{\ref{LAM}} 
\and H.~Lammer\inst{\ref{Graz}} 
\and C.~Lovis\inst{\ref{Geneve}}
\and P.~J.~MacQueen\inst{\ref{McDonald}}
\and T.~Mazeh\inst{\ref{Telaviv}}
\and M.~Ollivier\inst{\ref{IAS}} 
\and M.~P\"atzold\inst{\ref{Koeln}} 
\and D.~Queloz\inst{\ref{Geneve}}
\and H.~Rauer\inst{\ref{DLR},\ref{ZAA}} 
\and D.~Rouan\inst{\ref{LESIA}}
\and A.~Santerne\inst{\ref{LAM}} 
\and J.~Schneider\inst{\ref{LUTh}} 
\and B.~Tingley\inst{\ref{IAC},\ref{Laguna}} 
\and R.~Titz-Weider\inst{\ref{DLR}}
\and G.~Wuchterl\inst{\ref{Tautenburg}} 
}

%\and A.~Ofir\inst{\ref{Tel Aviv}}
%\and Th.~Pasternacki\inst{\ref{DLR}}

\institute{
Institute of Planetary Research, German Aerospace Center, Rutherfordstrasse 2, 12489 Berlin, Germany\label{DLR}
\and Laboratoire d'Astrophysique de Marseille, 38 rue Fr\'ed\'eric Joliot-Curie, 13388 Marseille cedex 13, France\label{LAM}
\and LUTH, Observatoire de Paris, UMR 8102 CNRS, Universit\'e Paris Diderot; 5 place Jules Janssen, 92195 Meudon, France\label{LUTh}
\and Research and Scientific Support Department, ESTEC/ESA, PO Box 299, 2200 AG Noordwijk, The Netherlands\label{ESA} 
\and Department of Physics, Denys Wilkinson Building Keble Road, Oxford, OX1 3RH, UK\label{Oxford}
\and Observatoire de l'Universit\'e de Gen\`eve, 51 chemin des Maillettes, 1290 Sauverny, Switzerland\label{Geneve}
\and LESIA, UMR 8109 CNRS, Observatoire de Paris, UPMC, Universit\'e Paris-Diderot, 5 place J. Janssen, 92195 Meudon, France\label{LESIA}
\and Institut d'Astrophysique Spatiale, Universit\'e Paris XI, F-91405 Orsay, France\label{IAS}
\and McDonald Observatory, University of Texas at Austin, Austin, 78712 TX, USA\label{McDonald}
\and Instituto de Astrof{\'i}sica de Canarias, E-38205 La Laguna, Tenerife, Spain\label{IAC}
\and Universidad de La Laguna, Dept. de Astrof\'\i sica, 38200 La Laguna,Tenerife, Spain\label{Laguna}
\and Observatoire de Haute Provence, 04670 Saint Michel l'Observatoire, France\label{OHP}
\and Institut d'Astrophysique de Paris, UMR 7095 CNRS, Universit\'e Pierre \& Marie Curie, 98bis boulevard Arago, 75014 Paris, France\label{IAP}
\and Rheinisches Institut f\"ur Umweltforschung an der Universit\"at zu K\"oln, Aachener Strasse 209, 50931, Germany\label{Koeln}
\and University of Vienna, Institute of Astronomy, T\"urkenschanzstr. 17, A-1180 Vienna, Austria\label{Wien}
\and IAG-Universidade de Sao Paulo, Brasil\label{Brasil}
\and Universit\'e de Nice-Sophia Antipolis, CNRS UMR 6202, Observatoire de la C\^ote d'Azur, BP 4229, 06304 Nice Cedex 4, France\label{OCA}
\and University of Li\`ege, All\'ee du 6 ao\^ut 17, Sart Tilman, Li\`ege 1, Belgium\label{Liege}
\and Th\"uringer Landessternwarte, Sternwarte 5, Tautenburg 5, D-07778 Tautenburg, Germany\label{Tautenburg}
\and Space Research Institute, Austrian Academy of Science, Schmiedlstr. 6, A-8042 Graz, Austria\label{Graz}
\and School of Physics and Astronomy, Raymond and Beverly Sackler Faculty of Exact Sciences, Tel Aviv University, Tel Aviv, Israel\label{Telaviv}
\and Center for Astronomy and Astrophysics, TU Berlin, Hardenbergstr. 36, 10623 Berlin, Germany\label{ZAA}
}
\date{Received 3 April, 2011; accepted 27 April, 2011}

%% traditional abstract format  
\abstract{We report on the discovery of a hot Jupiter-type exoplanet,
CoRoT-17b,  detected by the CoRoT satellite. It has a mass of 
$2.43\pm0.30$\Mjup~and a radius of $1.02\pm0.07$\Rjup, while its mean density
is $2.82\pm0.38$ g/cm$^3$. CoRoT-17b is in a circular orbit with a period of
$3.7681\pm0.0003$ days. The host star is an old ($10.7\pm1.0$ Gyr) main-sequence
star, which makes it an intriguing object for planetary evolution studies. The
planet's internal composition is not well constrained and can range from pure
H/He to one that can contain $\sim$380 earth masses of heavier elements.}


\titlerunning{CoRoT-17b}
\authorrunning{Csizmadia et al.}

\maketitle

%
%____________________________________________________________________________
\section{Introduction}
\label{sec:introduction}

Similar to eclipsing binaries, which offer the "Royal Road" to understanding the physics
and nature of stars (Russell 1932; Batten 2005), transiting exoplanets are key objects in
terms of understanding the formation, evolution and properties of planets. In Russell's
context, 'Royal Road' means that we have a way to obtain data that were previously
unavailable by other observational methods (Batten 2005). The transits and - if they are
observable - occultations of exoplanets provide unique possibilities to determine e.g.
their mass, radius and orbital characteristics, atmospheric composition  
as well as internal structure.
The consecutive transit and occultation observations allow us to study the orbital element
changes caused by either gravitational interaction with another body or by tidal and/or
magnetic interactions between the star and the planet.

CoRoT (Convection, Rotation and planetary Transit) is a 27 cm  diameter
space-telescope (Baglin 2007). The goals of the mission are to obtain long-term
photometric data sets for asteroseismological studies on relatively bright
pulsating stars and to search for new transiting exoplanets.  
Here we report the detection of another hot Jupiter, named CoRoT-17b. 
It orbits a faint star ($V \approx 15.5$) in a 3.76 days orbit. As we will show
below, the host star is evolved and quite old: its age is $10.7\pm1.0$ Gyr.

% table IDs, coordinates and magnitudes
\begin{table}
\caption{IDs, coordinates, and magnitudes of CoRoT-17.}
\centering
\begin{tabular}{lcc}       
\hline\hline                 
CoRoT window ID & LRc03\_E2\_2182  \\
CoRoT ID        & 311519570        \\
2MASS           & 2MASS18344782-0636440 \\
UCAC3           & 3UC 167-181321   \\
USNO-A2         & 0825-12387389    \\ 
USNO-B1         & 0833-042306      \\ 
GSC 2.2         & S3003-3212601    \\
GSC 2.3         & S9O9002601       \\
NOMAD1          & 0833-0454552     \\
\\
\multicolumn{2}{l}{Coordinates} \\
\hline            
RA (J2000)  &  $18$h $34$m $47.82$s         \\
Dec (J2000) &  $-6^\circ$ $36$' $44.04$'' \\
\\
\multicolumn{3}{l}{Magnitudes} \\
\hline
Filter & Mag \& Error & Source \\
\hline
r'     & 15.346$\pm$0.007 & ExoDat$^a$ \\
i'     & 14.521$\pm$0.011 &ExoDat \\
$V$   & 15.46       & ExoDat \\
$J$   & 13.174 $\pm$ 0.036 & 2MASS \\
$H$   & 12.615 $\pm$ 0.062 & 2MASS \\
$K$   & 12.472 $\pm$ 0.054 & 2MASS \\
$[3.6]$ & 12.232 $\pm$0.068 & IRAC \\
$[4.5]$ & 12.181 $\pm$0.069 & IRAC \\
$[5.8]$ & 12.149 $\pm$0.142 & IRAC \\
$[8.0]$ & 12.079 $\pm$0.151 & IRAC \\
\hline
\end{tabular}
\tablefoot{\tablefoottext{a}{Deleuil et al. (2009)}
}
\label{informationtable}      
\end{table}

%
%____________________________________________________________________________
\section{CoRoT observations and their analysis}
\label{sec:corot_lc}

\subsection{Data}
\label{subsec:corot_lc_data}

\begin{figure}%[t]
  \begin{center}
  \includegraphics[width=7.62cm]{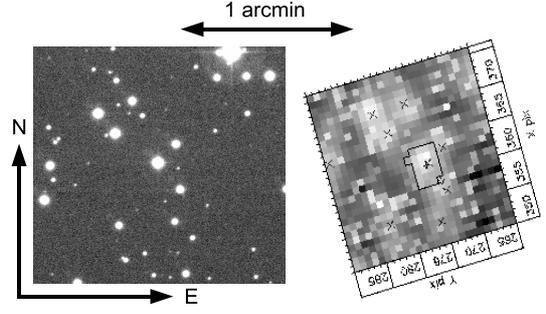}
  \end{center}
  \caption{Sky area around CoRoT-17 (star in the centre). Left: g-filter
image with a resolution of 0.7" taken with the CFHT telescope. Right:
image taken by CoRoT at the same scale and orientation. The jagged
outline in its centre is the photometric aperture mask; indicated are
also CoRoT's x and y image coordinates and positions of nearby stars
that are in the ExoDat database (Deleuil et al. 2009).
}
  \label{fig:findingchart}
\end{figure}

CoRoT-17 is located in the so-called LRc03 field, which was a medium-long field (for the
target's coordinates see Table~\ref{informationtable} and for the finding chart see
Fig.~\ref{fig:findingchart})\footnote{The interested reader can find more details about
the CoRoT observational strategy in Baglin et al. (2007), Boisnard et al. (2006) and 
Auvergne
et al. (2009).}.  
The star is listed in many photometric catalogues (for a selection see
Table~\ref{informationtable}), from which its spectral energy distribution is known
from the optical to mid-infrared.

The star was observed for 89.2 days, from HJD 2~454~925.4199 to HJD
2~455~014.6107,  corresponding to the period from 2009 April 03, 22 UT until 2009 July 02, 03 UT. In total, 14,861 data points were obtained in white light.  
However, some of these data points were obtained when the satellite crossed the
so-called South Atlantic Anomaly (SAA), which causes an increase in the high-energy
particle flux and strongly affects the observed fluxes. We excluded
these data points as well as those that were flagged by the pipeline for 
other reasons (Baudin et al 2008). For instance, we excluded all data points
that were acquired during the transition of the satellite from light to
Earth-shadow (see Auvergne et al. 2009) or vice versa (779 and 795 data points,
respectively). These
transitions cause considerable voltage and temperature fluctuations (Auvergne
2009). We kept 10,489 good data points (71.4\% of all the data points).  
The exposure time was uniformly 512 seconds for CoRoT-17b because of the
faintness of the star.

\begin{figure}%[t]
  \begin{center}
  \includegraphics[width=6cm,angle=270]{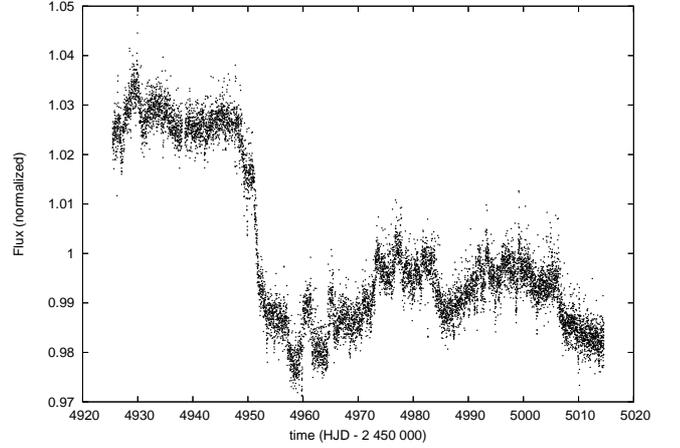}
  \end{center}
  \caption{
    Light curve of CoRoT-17 after removing the most obvious jumps.
  }
  \label{fig:fulllc}
\end{figure}

The original observations were reduced by the CoRoT pipeline. The jumps, caused
by high-energy particle impacts (cosmic ray events, see Auvergne et al 2009),
were corrected by us. The result can be seen in Figure~\ref{fig:fulllc}.

The light curve also shows some non-periodic modulation with an amplitude of 
$\sim2$\%, probably caused by stellar activity. The spectroscopic analysis of
the star (Sect. 3.3) yielded $v \sin i \approx 4.5\pm0.5$ km/s projected
rotational velocity and the combined spectroscopic and photometric analysis of
the stellar properties gave $R_{\mathrm{star}} = 1.59 R_\odot$ (Sect. 3.3). From
this one can expect $P = 20.1$ days rotational period for the star. We
calculated the Lomb-Scargle periodogram of the light curve, but no clear
periodicty nor any peak in the periodogram relatively close to this expected
value were found.

The transits were detected in the so-called 'alarm-mode' (Quentin et al. 
2006 Surace et al. 2008). Twenty-five transit events were observed by CoRoT with 
an orbital period of $3.7681$ days. These events are analysed in Sect. 
2.2. The duration of the transit was measured to be 4.72 hours and its 
depth is approximately 0.44\% (see Figure~\ref{fig:foldedlc} and 
Table~\ref{planparams}.).

\subsection{Transit modelling}
%\subsection{Data reduction}

\begin{figure}%[t]
  \begin{center}
  \includegraphics[width=7.62cm]{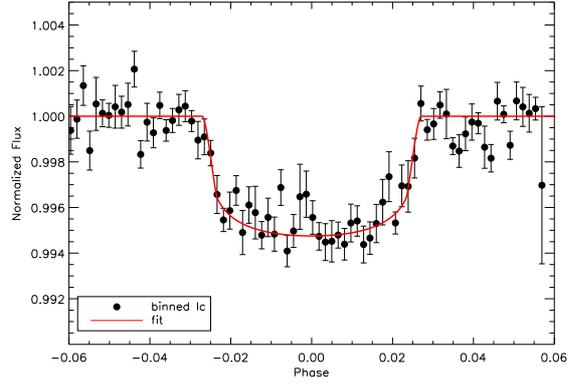}
  \end{center}
  \caption{
   Light curve solution of CoRoT-17. For the sake of clarity, we superimposed 
   the fit (red line) on the binned data instead of the original ones. The 
   features on the bootom of the light curve (sudden flux increases) maybe 
   caused by some spot crossing.  
  }
  \label{fig:foldedlc}
\end{figure}

Owing to the very complicated nature of the light curve, which is affected by various
instrumental effects such as jumps, SAA-crossing etc., we constructed the transit
light curve for the modelling in the following way. We excluded transits 5, 8,
12, 18, and 24, which were strongly altered by instrumental effects. Then we
folded the light curve as shown in Fig.~\ref{fig:foldedlc}. 

The instant of the centre of the $i$th transit is obviously $T_i = T_0 + iP$. 
Preceeding and following the beginning and the end of each transit, we cut one transit 
duration-long part of the light curve before and after the transit and fitted them by a 
parabola. Next, we selected the data points from the interval $T_i - 1.5D < t < T_i + 1.5D$ 
and then divided these light curve segments by the corresponding parabola. Here $D$ is the 
duration of the transit. After that we removed outliers by applying a 5$\sigma$ 
clipping. As a final step, we repeated this procedure again.

%\subsection{Transit modeling}

The transit light curve (Fig.~\ref{fig:foldedlc}) was fitted by a model to
obtain the relevant geometrical and physical parameters of the system. We used
the Mandel \& Agol (2002) model and a specified genetic algorithm (Geem et al.
2001) for the transit fit. Genetic algorithms have already successfully been
used to model the light curves of eclipsing binary stars, and it was also
applied to transit light curves (Fridlund et al 2010). Genetic algorithms have
the advantage that it is possible to map the whole parameter space, therefore  
it is more likely to find the global minimum in the parameter hyperspace. The
errors of the parameters can be easily estimated using a genetic algorithm
approach, too. We give the $1\sigma$-error bars, estimated from the width of the
distribution of points, which are between $\chi^2_{min}$ and $\chi^2_{min}+1$.

Our free parameters were: semi-major axis to the stellar radius ratio
($a/R_{star}$), planet-to-stellar radius ratio ($k$), impact parameter ($b=a\cos
i /R_\mathrm{star}$, where $i$ is the inclination), and the combination of $u_+
= u_1 + u_2$, while we kept fixed $u_- = u_1 - u_2$. Here $u_1$ and $u_2$ are
the linear and the quadratic limb-darkening coefficients, respectively. The
usage of these combinations was suggested by Brown et al (2001) and they are
widely used in the transit modelling studies. 

The contamination factor was a free parameter as well. This parameter gives an
estimate of the fraction of the total observed flux that comes from different
stars and not from the target object, because of the large PSF of CoRot
(Figure~\ref{fig:findingchart}). Bord\'e et al (2010) described in detail how to
determine this factor for CoRoT-targets. Applying the same method, we adopted $8
\pm 4$\% for the contamination factor and it could vary within these
constraints.

In total, we had five free parameters. Because the radial velocity curve indicates
a probably circular orbit, we assumed that the eccentricity is zero for the
light curve modelling .

We calculated 13 different models, each of which had a fixed,
but different value of $u_-$. The fixed values of this combination of limb
darkening coefficients were -0.3, -0.2, -0.1, ..., 0.7, 0.8 and 0.9. Convergence
was reached after $\sim$40~000 iterations at every fixed $u_-$ value.
The ``best``
light curve solution was then defined as the "centre of gravity"-like
average-value of the 1000 lowest $\chi^2$ at every fixed $u_-$ value.

The $\chi^2$-values showed an absolute minimum value at $u_- = 0.2$ where $u_+$
was equal to $0.65 \pm 0.17$. We checked whether our limb-darkening coefficients
agree with the theoretical ones, derived from the effective temperature, $logg$
and metallicity of the star (see Sect. 3.3). Using the tables of Sing (2010), we
found that theoretically they are $u_- = 0.23 \pm 0.03$ and  $u_+ = 0.68 \pm 0.03$ 
for CoRoT-17. (The uncertainties of the theoretical values stem from the uncertainties 
of the stellar parameters.) The modelled values are close to this prediction. We do 
not investigate this problem further, because the photometric 
signal-to-noise ratio is 
not sufficient to determine both limb-darkening coefficients simultaneously, and 
because  
the 
found best value was within the error bars of the predicted one, 
the agreement is satisfactory. The finally adopted solution is shown in 
Fig.~\ref{fig:foldedlc}. The results of the transit light curve modelling is given in 
Table~\ref{planparams}. In the subsequent analysis we need the $M_{star}^{1/3} / R_{star}$ 
for the determination of the stellar mass and radius. Re-arranging Kepler's third law 
(Roberts 1899; Winn 2010), we calculate a value for this quantity of $0.61\pm0.03$ 
(solar units).

%_____________________________________________________________________________
\subsection{Parameter correlations}

The result of transit lightcurve modelling can be degenerate, leading to 
equally acceptable solutions. For CoRoT-3b, for instance, the light curve
modelling itself showed that in the inclination - limb darkening coefficient
plane there are two solutions, which cannot be distinguished from each other
(Deleuil et al. 2008). It was already noticed in Brown et al. (2001) that at
low impact parameters the solutions are degenerate in the inclination-limb
darkening coefficient plane if the observational errors are relatively large.
Because CoRoT-17 is relatively faint, the light curve modelling most likely forms 
an ill-posed problem because of the noise levels involved. Therefore it is 
important to investigate how the various derived parameters are correlated.

\begin{figure*}
   \centering
   \includegraphics[width=7.2cm]{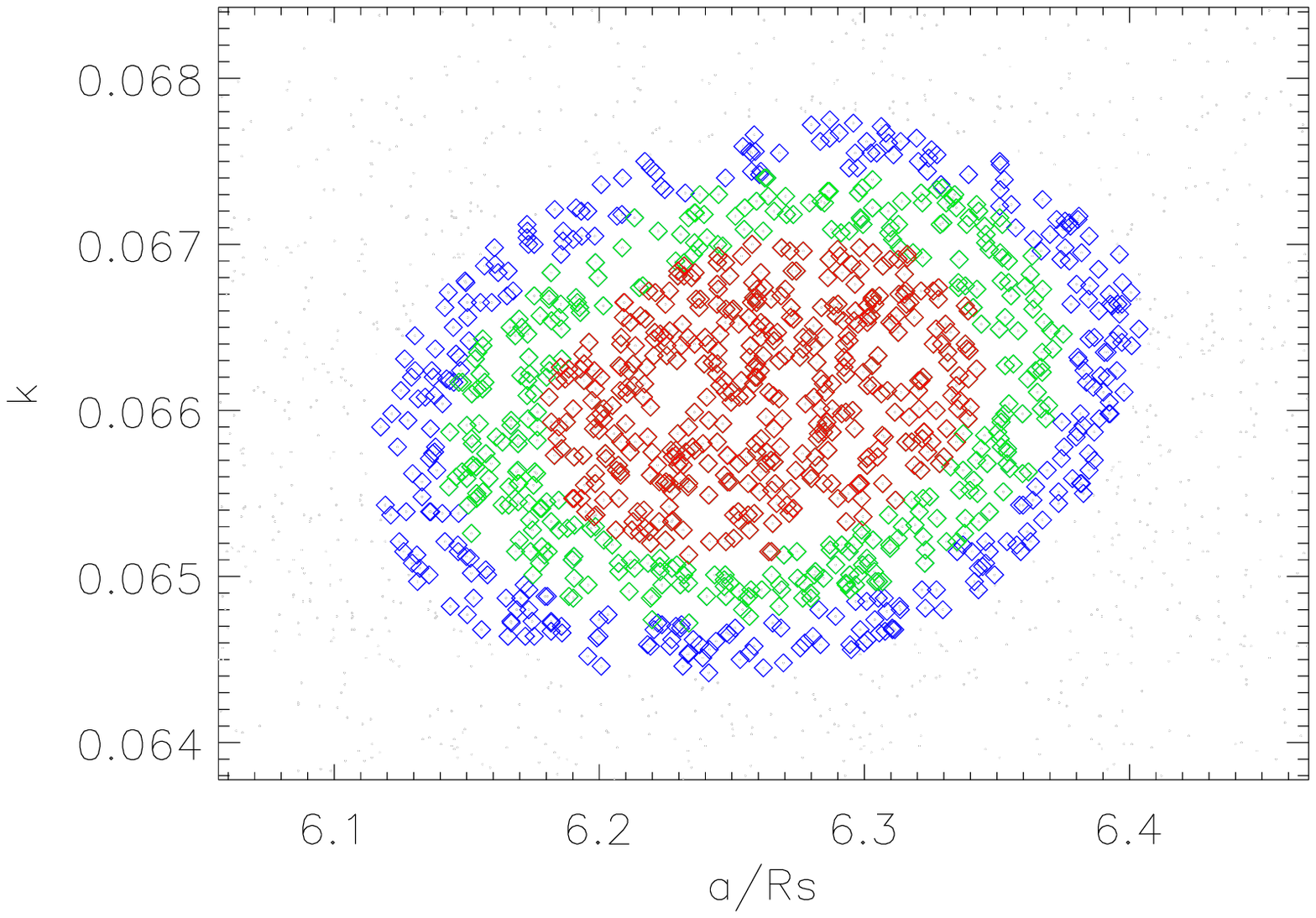}
   \includegraphics[width=7.2cm]{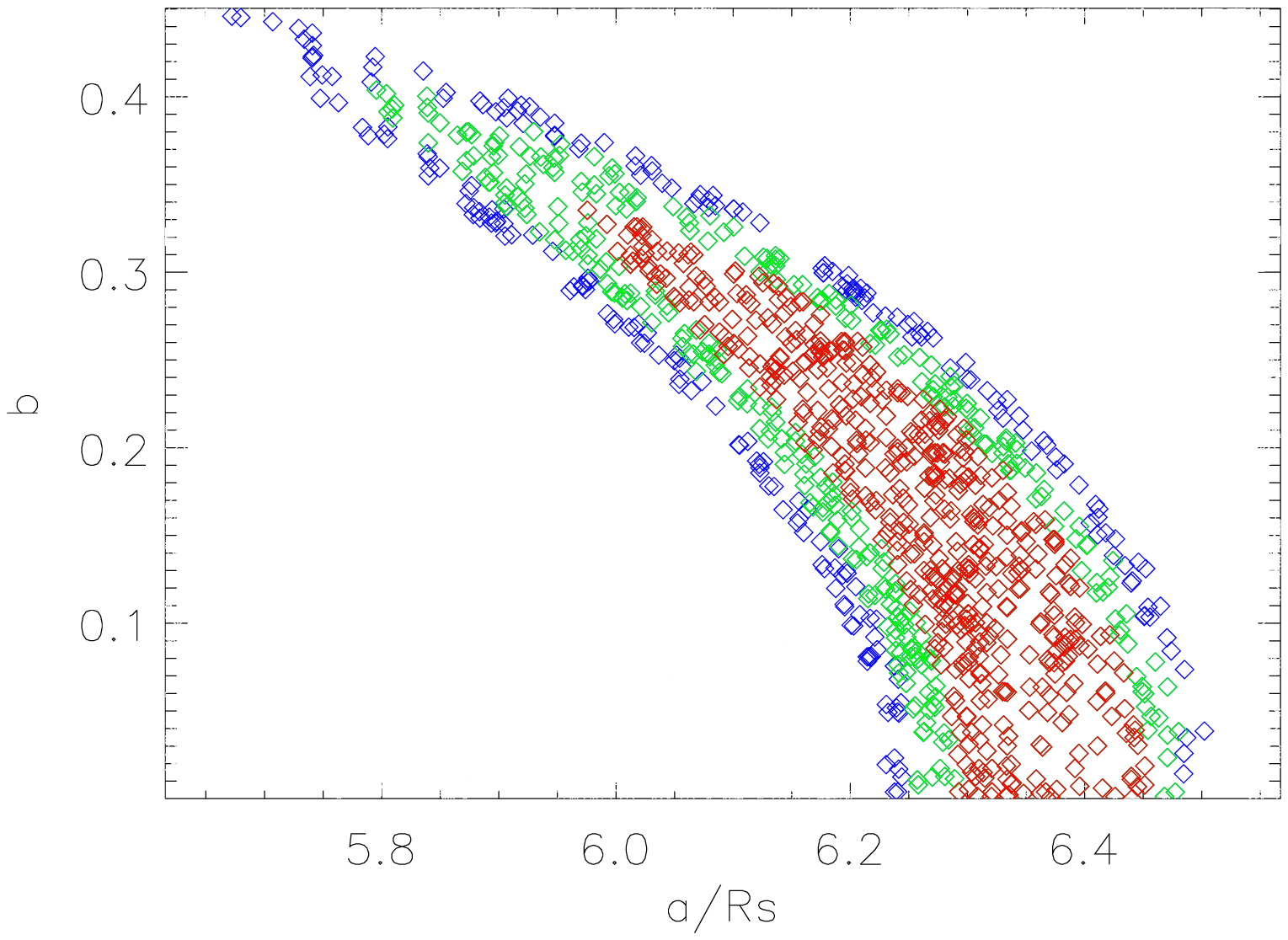}
   \includegraphics[width=7.2cm]{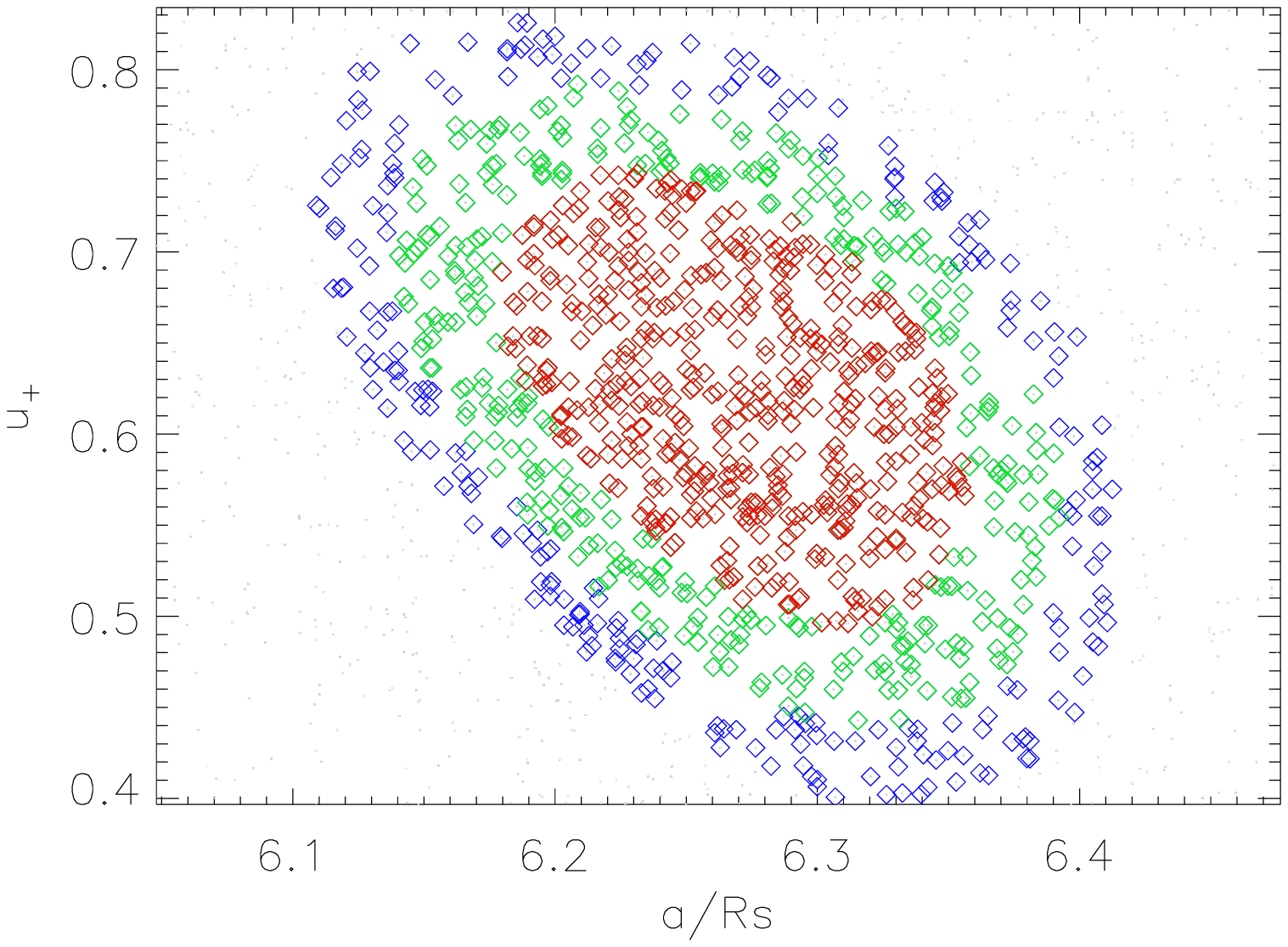}
   \includegraphics[width=7.2cm]{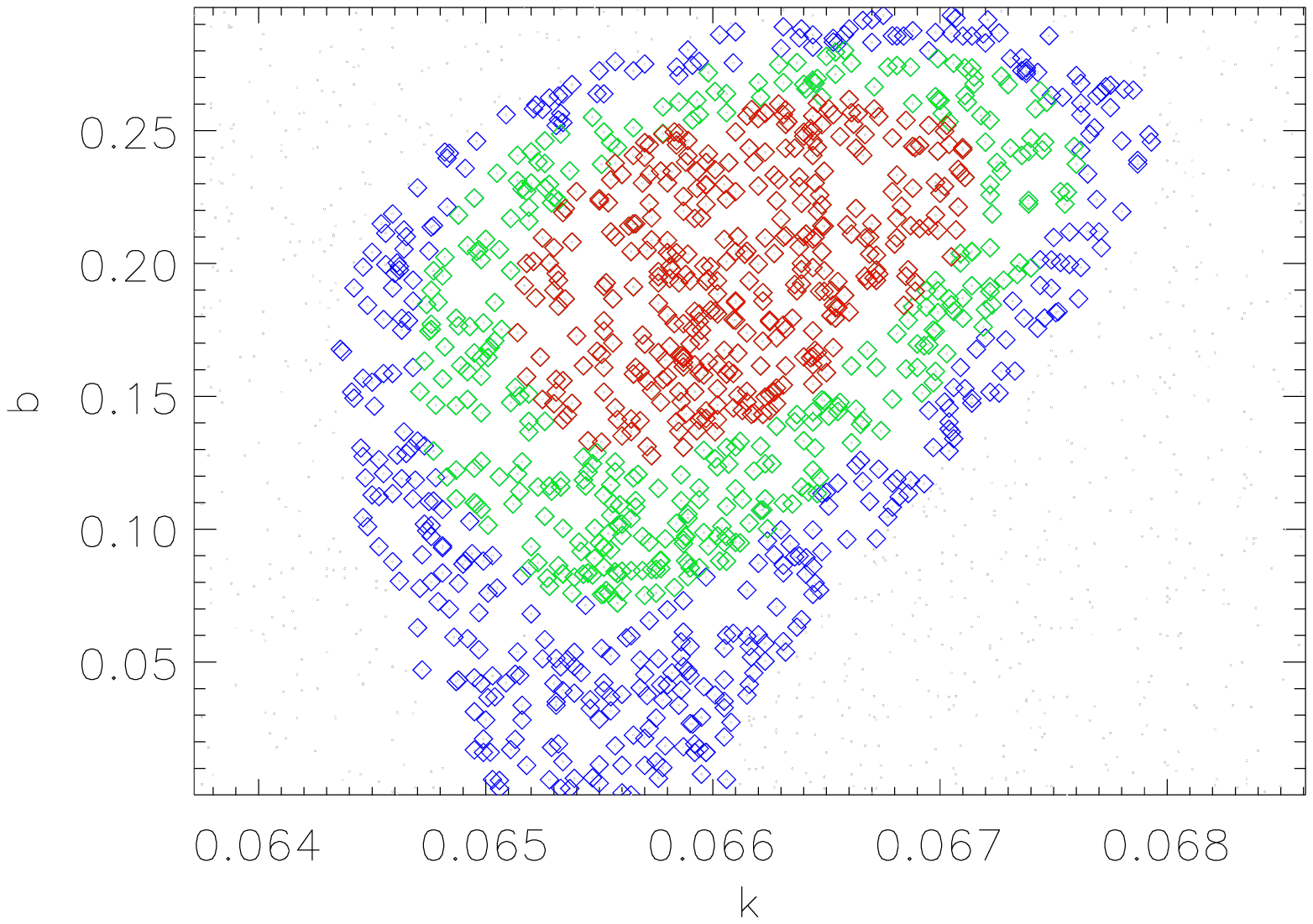}
   \includegraphics[width=7.2cm]{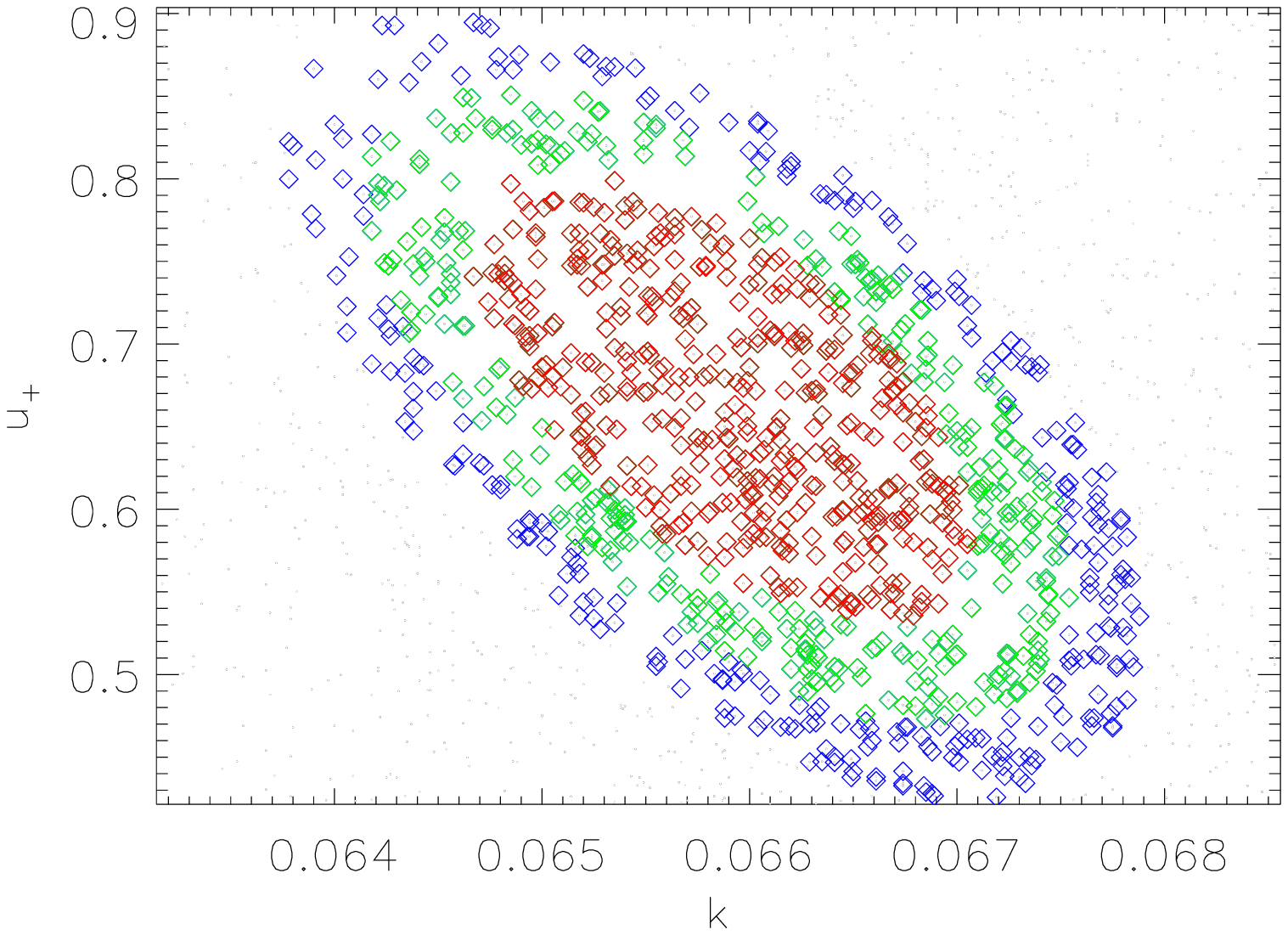}
   \includegraphics[width=7.2cm]{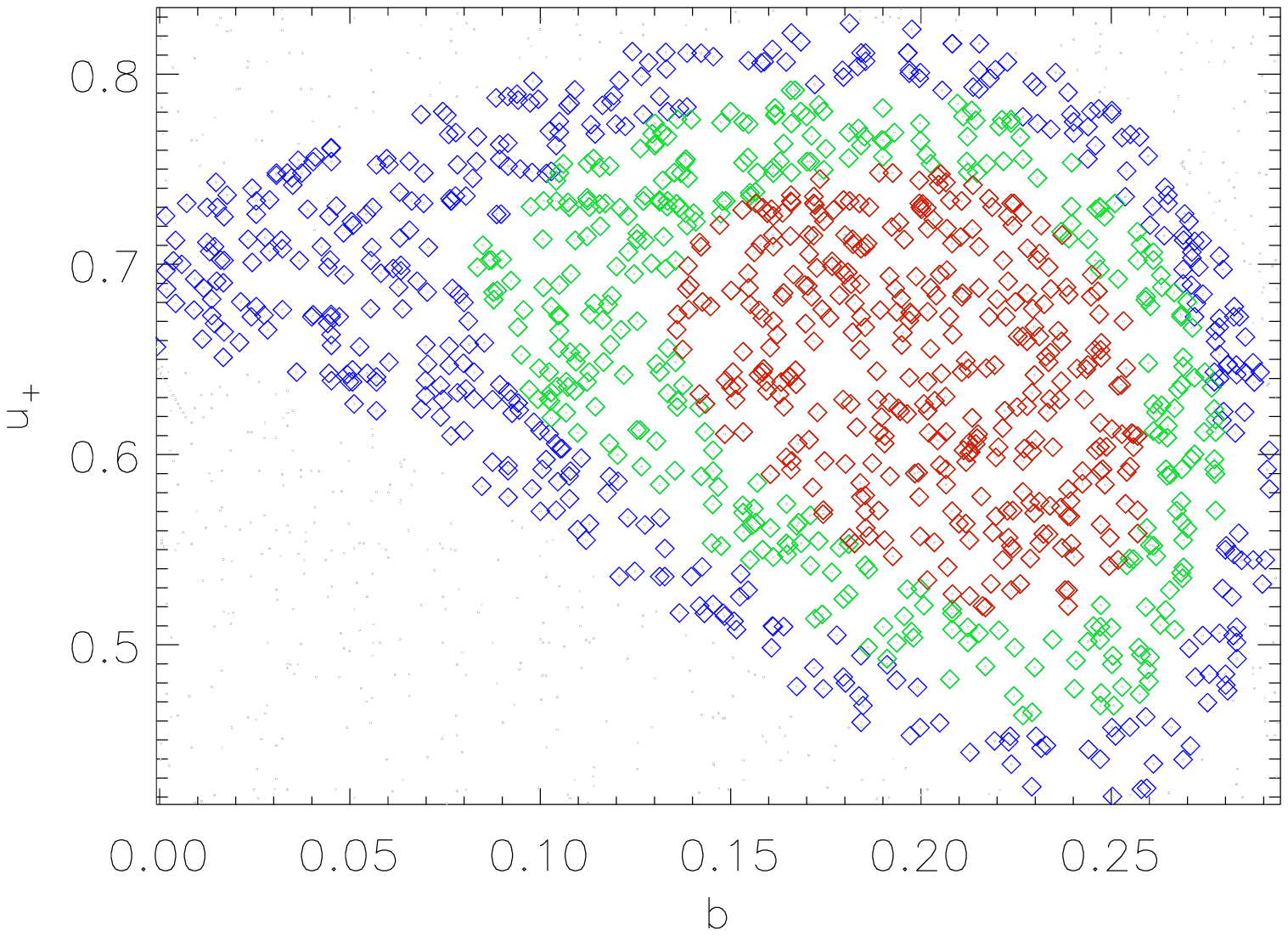}
   \caption{Parameter-correlation diagrams for CoRoT-17.}
   \label{fig:figcorrs}
   \end{figure*}

To do that, we selected six combinations of four free parameters ($a/R_{star}$,
$k$, $b$ and $u_+$, see Fig.~\ref{fig:figcorrs}). Each of these combinations
consisted of two parameters. Then we fixed the other parameters at their values
given in Table~\ref{planparams} determined before, while several hundreds of
randomly selected combinations of the two investigated parameters were varied
over a large interval of values. The panels of Fig.~\ref{fig:figcorrs} show 
the models within 1$\sigma$ (red), 2$\sigma$ (green) and 3$\sigma$ (blue). 

We found that $a/R_{star}$ and $k$ are not correlated, while we found negligible 
correlation for the pair of  
$a/R_{star}$ -- $u_+$, and the correlation between $k$ -- $u_+$ is moderate. It is 
remarkable that the $k$ -- $b$ 
and $b$ -- $u_+$ diagram has a non-symmetrical distribution. The well-known 
correlation between $a/R_{star}$ and the impact parameter $b$ can also be seen. 
From the top right panel of Fig.~\ref{fig:figcorrs} we can only give wide limits
for these two parameters.

Studying these correlation diagrams, we concluded that all the parameters are
well determined and unique, but the semi-major axis and the impact parameter are
correlated. An impact parameter between $0$ and $0.3$ is the most
probable value, but we are not able to constrain it better from the presently
available photometry. It is safer to state that the semi-major axis/stellar
radius ratio is most probable between $6.0$--$6.4$, but it is also hard to give
better constraints. To gain more precise values one requires even more precise
photometry for this faint star.

\begin{table}
\caption{Physical and geometrical parameters of the CoRoT-17 system. Inclination ($i$) was calculated from the $a/R_\mathrm{star}$ ratio and from the impact parameter $b$. The parameter $M^{1/3} / R$ can be calculated from the orbital period and from the $a/R_{star}$ value (see e.g. Winn 2010). $T_{eq}$ is the equlibirium temperature of the planet.}
\begin{tabular}{ll}
\hline
\multicolumn{2}{l}{Determined from photometry} \\
\hline
Epoch of periastron $T_{0}$ [HJD-2400000] & 54~923.3093$\pm$0.0036 \\
Orbital period (days) & 3.7681$\pm$0.0003\\
Duration of the transit (hours) & 4.72 \\
Depth of the transit (\%) & 0.44 \\
\hline
\multicolumn{2}{l}{Determined from RV-measurements} \\
\hline
Orbital eccentricity $e$  & 0 [fixed] \\
Argument of periastron $\omega$ [deg] & 0 [fixed] \\ 
Radial velocity semi-amplitude $K$ [\ms] & 312.4 $\pm$ 29.0 \\
Systemic velocity  $V_{\gamma}$ [\kms] & 54.770 $\pm$ 0.008 \\
O-C residuals [\ms] & 77 \\
\hline
\multicolumn{2}{l}{Determined from spectral analysis of the star} \\
\hline
$T_{eff}$  [K]               & 5740$\pm$80   \\
$\log g_\mathrm{star}$ [cgs] & 4.40$\pm$0.10 \\
$[Fe/H]$                     & 0.0$\pm$0.1   \\
$v \sin i$ [\kms]            & 4.5$\pm$0.5\\
Spectral type                & G2V \\
\hline
\multicolumn{2}{l}{Determined from light curve modelling} \\
$a/R_\mathrm{star}$ & 6.23$\pm$0.24\\
$b$               & 0.18$\pm$0.16\\
$i$ [deg]         & 88.34$\pm$1.54\\
$k$               & 0.0661$\pm$0.0019\\
$u_+$             & 0.65$\pm$0.17\\
$u_-$             & 0.2 (fixed)\\
\hline
\multicolumn{2}{l}{Combined results} \\
\hline
Stellar mean density $\rho_{st}$ [solar] & 0.23 $\pm$ 0.02 \\
Stellar mass $M_{st}$ [solar] & 1.04 $\pm$ 0.10 \\
Stellar radius $R_{st}$ [solar] & 1.59 $\pm$ 0.07 \\
Stellar age [Gyrs]              & 10.7$\pm$1.0 \\
Orbital semi-major axis $a$ [AU] & 0.0461 $\pm$ 0.0008 \\
Planet mass   $M_{p}$ [M$_J$ ] &  2.43 $\pm$ 0.30 \\
Planet radius $R_{p}$ [R$_J$ ] &  1.02 $\pm$ 0.07 \\
Planet mean density $\rho_{p}$ [$gcm^-3$] &  2.82 $\pm$ 0.38 \\
$M_{st}^{1/3} / R_{st}$ [solar units]  &0.61 $\pm$ 0.03 \\
$T_{eq}$ [K] & 1626 $\pm$ 31 \\
Distance [pc] & 920 $\pm$ 50 \\
$A_V$   [mag] & 2.60$\pm$ 0.10  \\
\hline
\hline
\end{tabular}
\label{planparams}
\end{table}

\begin{table*}
\caption{Telescopes used to assess the transit events observed by CoRoT and
the log of observations. See text for discussion. For TRAPPIST see Gillon et al.
(2011) while for BEST II see Erikson et al. (2011).}
\begin{tabular}{lll}
\hline
Telescope &	Date of observation &	Note \\
\hline
CFHT/MEGACAM              & 2010 May 17 & ON/OFF \\
                          & 2010 June 5 & ON/OFF \\
OGS (Tenerife)            & 2010 June 9 & ON \\
                          & 2010 June 10& OFF \\
TRAPPIST (La Silla)       & 2010 May 24 & Only egress phase was observed\\
BEST II (Cerro Armazones) & 2010 June 5 & Photometry during the full event \\
\hline
\hline
\end{tabular}
\label{telescopes}
\end{table*}

%____________________________________________________________________________
\section{Ground-based observations and their analysis}
\label{sec:ground_observations}

\subsection{Photometric measurements}
\label{subsec:photometric_measurements}

As part of the ground-based photometric follow-up programme of CoRoT candidates,
observations of CoRoT-17 were undertaken from several telescopes. The aim of
these observations was to assess if the transit-events observed by CoRoT
really arise from the CoRoT target star and not from any contaminating nearby
variable source. For more details on this follow-up programme, its motivation,
techniques and results, see Deeg et al. (2009).  

The log of observations and the list of the used telescopes can be found in
Table~\ref{telescopes}. Of these observations, none showed any sign of relevant
variability on any of several stars that are contaminating the CoRoT-aperture
(Fig.~\ref{fig:findingchart}). These stars would have to display eclipses with
amplitudes of at least 7\%, which would have been detected by any of the
ground-observations that were undertaken. Only the second night from CFHT on
2010 June 5 also showed a significant variation on the target itself with an
amplitude of 0.55\%, close to the expected one from CoRoT. The absence of a
signal from the target in the other data has the following reasons: (i) the
observations obtained in May suffered from a preliminary ephemeris and thus
they did not see the deepest part of the transit. As a result of this
experience, the ephemeris was refined and replaced in early June, 2010. (ii) The
transit event in June was too shallow for OGS and BEST II and that is why these
data were used only to exclude possible background eclipsing binaries.  

In conclusion, the absence of variability in the brighter contaminating stars as
well as the detection of a correct brightness-variation on the target assured
that the signal observed by CoRoT really arose from the target.

%............................................................................
\subsection{Radial velocity measurements and orbit of CoRoT-17b}

HARPS is used to establish the planetary nature of CoRoT candidates and to
identify binaries, in coordination with SOPHIE at Observatoire de Haute-Provence
and HIRES at the Keck telescope (Bouchy, Moutou and Queloz, 2009). 
For radial velocity measurements of CoRoT-17, we used only
HARPS (Mayor et al 2003). RV positions are estimated by cross-correlation of the
stellar spectrum with a numerical mask (Baranne et al 1996, Pepe et al 2003). It
commonly reaches uncertainties of 1 m/s on bright stars, up to typically 50-100
m/s for a V$\simeq 15.5$ as CoRoT-17. 

Between May and August, 2010, CoRoT-17 has been observed with HARPS at a spectral
resolution of 110,000, together with a simultaneous on-sky reference in fiber B for
monitoring the sky background. In total, seventeen measurements have been secured. Some
measurements have been acquired when the moon was up, but the velocity difference between the
stellar spectrum and the barycentric Earth RV is always larger than 22 km/s, so that the
stellar cross-correlation function is not significantly contaminated. Hence, the RV
measurements have not been corrected for, even when the sky background is detected. With
an exposure time  of 1 hour, a typical signal-to-noise ratio of 8 is achieved on this
target.  The average uncertainty of the derived velocities is 74 m/s (see Table 
\ref{rvtab}).

\begin{figure}[h]
\centering
\includegraphics[width=8.5cm]{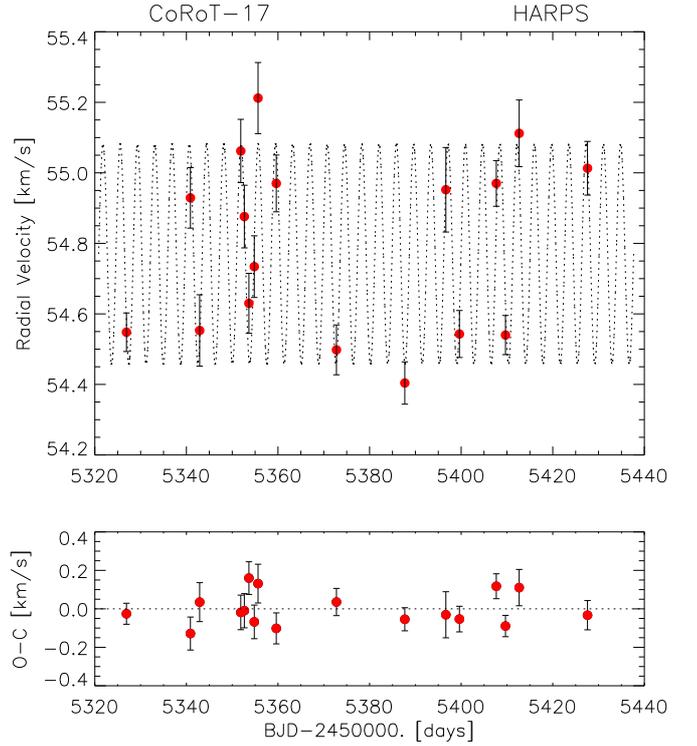}
\caption{Radial-velocity variations of CoRoT-17 obtained with HARPS as a 
function of time. The Keplerian best fit is superimposed. The bottom plot 
shows the residuals to this model.}
\label{rv1}
\end{figure}

\begin{figure}[h]
\centering
\includegraphics[width=8.5cm]{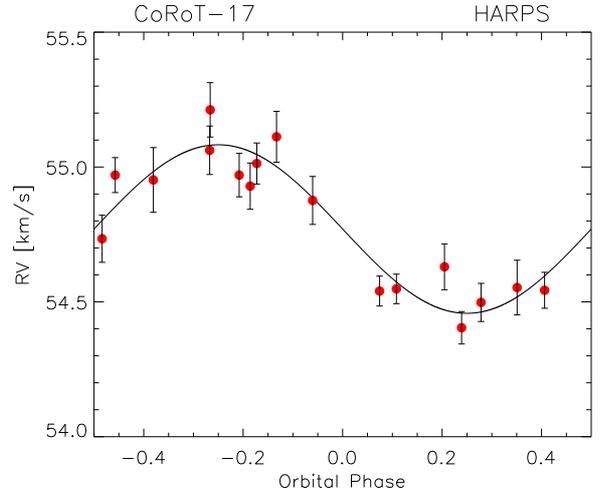}
\caption{RV data of CoRoT-17 folded to the ephemeris derived from 
CoRoT transits.}
\label{rv2}
\end{figure}

The $rms$ of the RV time series is 244 m/s, indicating a fluctuation three times 
higher than the error bars (Fig. \ref{rv1}). When phased at the CoRoT period and 
transit ephemeris, a sinusoidal variation is clearly observed (Fig. \ref{rv2}), 
characterized by a semi-amplitude $K =$ 312 m/s and systemic velocity 
$V_0 = $54.8 km/s. The cross-correlation function is narrow ($FWHM=9.6$ km/s), 
typical of a main-sequence slow-rotating star. The estimated projected velocity 
of the star is 5 $\pm$ 1 km/s (this value is slightly refined in the next paragraph, using the co-added spectra). The Keplerian solution is shown in Figures \ref{rv1} and \ref{rv2}, where the orbital eccentricity is fixed to 0. Assuming $e$ is a free parameter, a value of 0.08 $\pm$ 0.12 is found. Our current data set does not significantly constrain the eccentricity. The final $rms$ of the residuals is 77 m/s, a value close to individual error bars; the achieved 
reduced $\chi^2$ is 1.25 when period, time, and eccentricity are fixed. 

\begin{figure}[h]
\centering
\includegraphics[width=8.5cm]{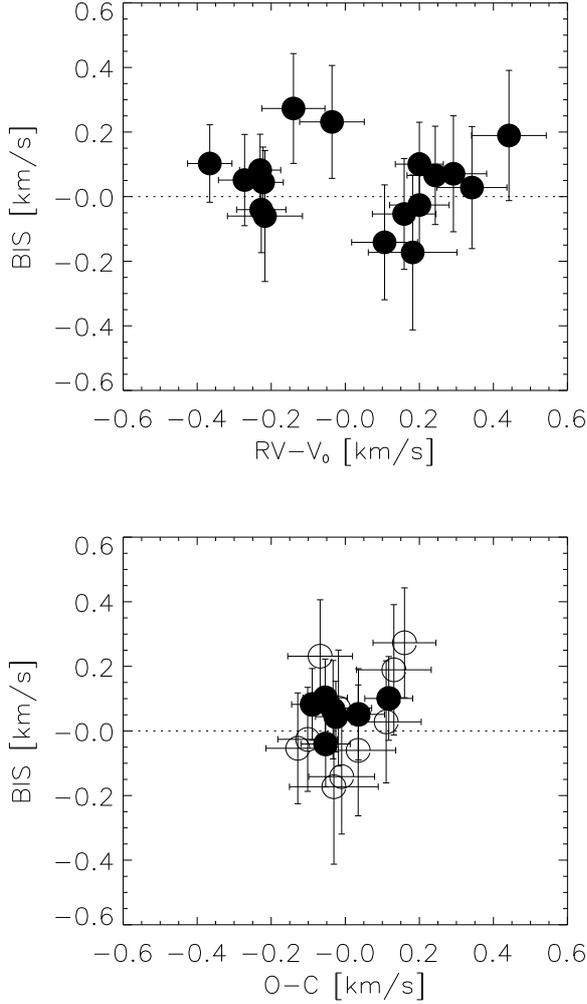}
\caption{Variations of the bisector span of CoRoT-17 RV data as a 
function of the measured velocities (top) and the residuals to the Keplerian 
fit (bottom). The first plot shows no correlation, a good indication for the 
planetary hypothesis as opposed to a blend binary signal. The second plot shows a weak 
correlation, owing the lowest SNR data of the sample (open points 
show the data where SNR is less than 5). }
\label{bis}
\end{figure}

Common tests on the bisector behaviour have been performed, as shown in 
Figure \ref{bis}. The bisector span is estimated between the wings and the 
core of the cross-correlation function, and its behaviour as a function of 
the radial velocities and the residuals to the Keplerian fit are examined. 

No trend is found between the bisector span and the velocity, with a correlation 
coefficient of 7\%; this indicates a low probability of a triple system or a 
diluted binary scenario. A slight trend (correlation of 41\%) is observed 
between the bisector slope and the residuals to the fit. However, as suggested  
in Figure \ref{bis}, this bisector behaviour is dominated by the data that 
have the lowest signal-to-noise ratios. This test shows the limitation of the 
bisector diagnostics, when the wings of the cross-correlation function are 
strongly affected by the spectrum continuum, at SNR typically less than 
$\sim$5. Because the bisector has no detected variation in phase with the 
radial-velocity variations, we conclude that the scenario involving a 
transiting planetary companion of CoRoT-17 is the only explanation for the observed radial velocity variations.

%............................................................................
\subsection{Spectral analysis of the host star}
\label{subsec:spectroscopic_measurements}

As part of the NASA's key science programme in support of the CoRoT mission, 
CoRoT-17 was also observed on 2010 June 20 (UT) with the HIRES spectrograph 
(Vogt et al. 1994) mounted on the Keck~I 10\,m telescope, at the Keck Observatory 
(Mauna Kea, Hawai'i). The red cross-disperser along with the $0\farcs861$ wide 
slit and 14$\arcsec$ tall decker were employed to properly subtract the sky 
background, yielding a resolving power of $R\approx50\,000$ and a wavelength 
coverage $3800 \leq \lambda \leq 7975$~\AA. Six consecutive spectra of 1200 
sec each were acquired to increase the S/N ratio and remove cosmic ray hits. 
The spectra were extracted and co-added with standard IRAF routines, giving a 
final S/N ratio of about 65 at 6000~\AA.

Following the standard practice already adopted in previous CoRoT discovery
papers (e.g., Deleuil et al. 2008; Fridlund et al. 2010; Gandolfi et al. 2010),
we used a HIRES co-added spectrum to derive
the fundamental physical parameters of CoRoT-17, i.e., effective temperature
($T_\mathrm{eff}$), surface gravity (log\,$g$), metallicity ($[\rm{Fe/H}]$), and
projected rotational velocity (\vsini). Some methods consist of comparing the
observed spectra with a grid of model atmospheres from \citet{Castelli04},
\citet{Coelho05} and \citet{Gustafsson08}, using spectral lines that are
sensitive to the different photospheric parameters. Other methods rely on the
use of spectral analysis packages like SME~2.1 \citep{Valenti96,Valenti05} and
VWA \citep{Bruntt04,Bruntt08,Bruntt10}. We found consistent results regardless
of the spectrum and procedure. The final values adopted for the above
mentioned physical parameters are $T_\mathrm{eff}= 5740\pm80$~K,
log\,$g=4.40\pm0.10$, $[\rm{Fe/H}]=0.0\pm0.1$, and $\vsini=4.5\pm0.5$~km/s
(Table~\ref{planparams}).

Using the $M_{star}^{1/3} / R_{star}$ value obtained from the transit light
curve modelling, we obtained the mass, radius, and age of the host star
via the same procedure we applied in earlier CoRoT planet-detection publications
(see e.g. Deleuil et al. 2008). 
The results are reported in Table~\ref{planparams}. Note 
that the star is quite evolved from the ZAMS, and based on the presently
available constraints (effective temperature, metallicity, $M_{star}^{1/3} /
R_{star}$ parameter), we found the age of the star to be 
$10.7\pm1.0$ Gyr.

\begin{table}
\caption{HARPS radial-velocity measurements of CoRoT-17}
\begin{tabular}{llll}
hline
BJD-24500000. & RV & $\sigma_{RV}$ & BIS \\
(days)  &[km/s] &[km/s] &[km/s]\\
\hline
55326.90632&	54.548&	0.055&		0.044	\\
55340.87074&	54.929&	0.086&		-0.054	\\
55342.89312&	54.553&	0.101&		-0.060	\\
55351.86818&	55.062&	0.090&		0.071	\\
55352.64808&	54.876&	0.089&		-0.141	\\
55353.64740&	54.630&	0.085&		0.273   \\
55354.82094&	54.734&	0.087&		0.231	\\
55355.64015&	55.212&	0.101&		0.189	\\
55359.62782&	54.970&	0.081&		-0.026	\\
55372.76478&	54.498&	0.071&		0.051	\\
55387.68993&	54.404&	0.060&		0.103	\\
55396.65855&	54.952&	0.120&		-0.173  \\
55399.62341&	54.543&	0.067&		-0.040	\\
55407.67391&	54.970&	0.065&		0.101	\\
55409.67746&	54.540&	0.056&		0.082   \\
55412.66468&	55.112&	0.095&		0.028 	\\
55427.58796&	55.013&	0.076&		0.065	\\
\hline
\end{tabular}
\label{rvtab}
\end{table}

%% RV FIGURES %%

%------------------------------------------------------------------------------
\subsection{The spectral energy distribution of the host star}

\begin{figure}
   \centering
   \includegraphics[width=9cm]{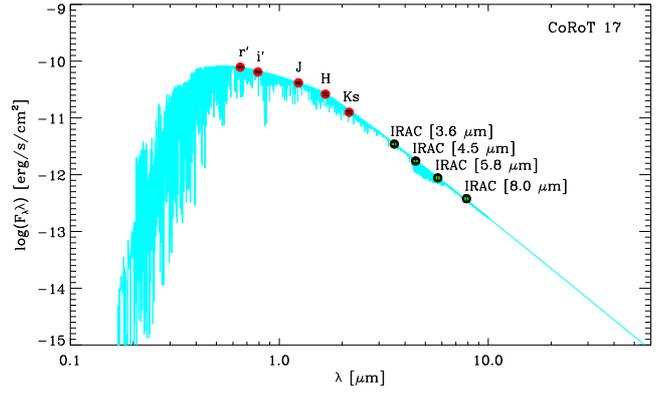}
   \caption{Spectral energy distribution of CoRoT-17. Dereddened  $r^\prime i^\prime JHKs$  and IRAC infrared data are represented with filled red and green dots, respectively. The model spectrum by Hauschildt et al. (1999) with the same temperature, radius, and metallicity as CoRoT-17 are plotted with a light-blue line.}
   \label{fig:sed}
  \end{figure}

The interstellar extinction ($A_{\mathrm V}$) and distance ($d$) to the star
 were derived using the $r^\prime i^\prime JHKs$ photometry from the ExoDat
 database combined with IRAC@Spitzer 3.6, 4.5, 5.8, and 8.0 $\mu m$ magnitudes
 (Table~\ref{informationtable}). The Spitzer-magnitudes were extracted by applying
 aperture photometry on the IRAC image cut-outs centred around CoRoT-17, as
 retrieved from the IRSA-NASA/IPAC Infrared Science
 Archive\footnote{http://irsa.ipac.caltech.edu/}. Following the method described
 in Gandolfi et al. (2008), we simultaneously fitted the colours ecompassed by
 the spectral energy distribution (SED, see Figure~\ref{fig:sed}) with synthetic
 magnitudes derived from the \emph{NextGen} stellar atmosphere model (Hauschildt
 et al. 1999) with the same $T_\mathrm{eff}$, log\,$g$, and $[\rm{Fe/H}]$ as the
 star. Assuming a normal total-to-selective extinction coefficient ratio
 $R_{\mathrm V}=3.1$ and a black body emission at the star's effective
 temperature and radius, we found $A_{\mathrm V}=2.60\pm 0.10$~mag and $d=920
 \pm 50$~pc. The IRAC 5.8 and 8.0 $\mu m$ images reveal a patchy nebula
 spatially projected towards the direction of CoRoT-17.
 The presence of foreground interstellar medium for
 such a low galactic latitude star ($b\approx0.7\degr$) might account for the
 derived high extinction value.

\begin{figure}[t]
  \centering
  \begin{minipage}[t]{0.49\textwidth}
      \includegraphics[width=\textwidth]{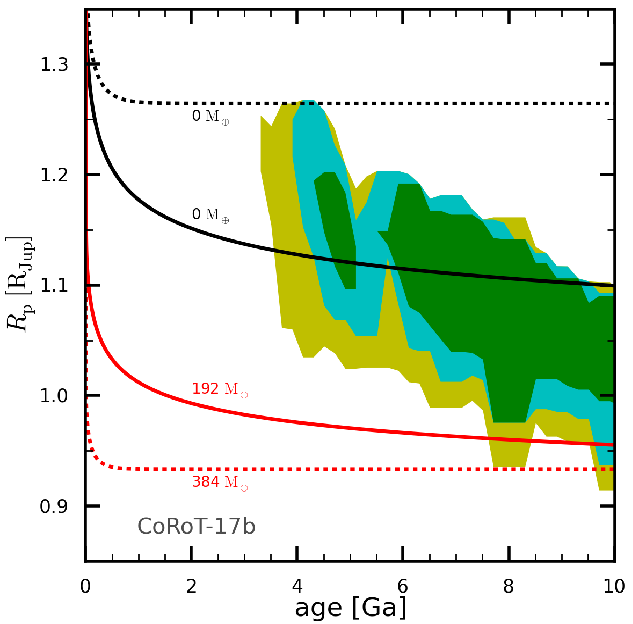}
  \end{minipage}
  \caption{Evolution of the size of CoRoT-17b (radius in Jupiter units) as a
function of age (in billion years), compared to constraints inferred from
CoRoT photometry, spectroscopy, radial velocimetry, and the PADOVA stellar 
isochrones. Green, blue, and yellow-green plain regions correspond to the
planetary radii and ages that result from stellar evolution models that match the 
inferred $\rho_{star}$ - $T_{eff}$ uncertainty ellipse within 1, 2, and
3$\sigma$, respectively. Planetary evolution models for a planet with a solar-composition
envelope and no dissipation are shown as plain lines and labelled according
to the value of the core mass (in earth masses); models in which 0.25\% of the
incoming flux is deposited at the planet's centre are shown as dotted lines.
Models with no core are shown in black, and with a core in red. These models
assume a total mass of 2.43\Mjup and an equilibrium temperature of 1626 K.}
  \label{fig:evolution}
\end{figure}

%
%____________________________________________________________________________
\section{Discussion}
\label{sec:discussion}

We reported the discovery of a transiting exoplanet around the faint
star CoRoT-17, detected by the CoRoT satellite. Extensive radial velocity,
spectroscopic and photometric follow-up measurements proved that the cause of
these transits is a transiting exoplanet that orbits the star. The measurements also 
helped to
determine the basic characteristics of this planetary system. The planet has a
mass of $2.43\pm0.30$\Mjup~and a radius of  $1.02\pm0.07$\Mjup, while the mean
density is $2.82\pm0.38$ g/cm$^3$ (Table~\ref{planparams}). The transit is
almost central and it has 0.44\% depth, but the CoRoT light curve is
contaminated by another star, which contributes $8\pm4$\% to the total
observed flux. The host star is an evolved G2V type star, with mass 
$1.04\pm0.10M_\odot$ and radius $1.59\pm0.07R_\odot$. The isochrone fit
yielded the age of the star, which is $10.7\pm1.0$ Gyr. This makes the system
one of the oldest known systems among the transiting exoplanets. 

CoRoT-17b is located in a quite common place on the mass -- radius diagram of
the known transiting exoplanets (see eg. \citealp{south10, swift10}). 
To probe
the possible composition of CoRoT-17b, we computed stellar and planetary
evolution models using PADOVA (Marigo et al. 2008; Bertelli et al. 2008) and CEPAM 
(Guillot \&
Morel 1995), as described in Guillot \& Havel (2011). The results are shown in
Fig~\ref{fig:evolution} where the evolution of the size of CoRoT-17b is plotted
as a function of the system age. The colours indicate the distance in standard
deviations from the inferred effective temperature and mean stellar density,
i.e., less than $1\sigma$ (green), $2\sigma$ (blue) and $3\sigma$ (yellow).
These constraints are compared to planetary evolution models for a homogeneous
solar-composition hydrogen-planet, with different hypotheses: (plain lines)
using a 'standard model', i.e., whithout additional sources of heat; (dotted
lines) by adding a fraction (0.25\%) of the incoming stellar energy and
dissipating it at the centre. These two cases correspond to standard recipes
used to explain the inflated giant exoplanets (Guillot 2008). In this context,
it is hard to constrain the mass of heavy elements needed to match the observed
planetary radius. The core mass can range from 0 (pure H-He planet) to $\sim$380
earth-masses ($\sim48$\% of the total mass), depending on the hypothesis we consider.
However, this diagram is not sensitive for the properties of CoRoT-17b, because
- depending on the assumptions - we can easily construct models with 192 and 384
$M_{earth}$ of heavy elements inside the planet, too, which are practically
undistinguishable at the age of the system, because at very late stages of
planetary evolution these two evolutionary tracks are closer to each other than
the uncertainty in the planetary radius.

Similarly to other known old planetary systems (e.g. HAT-P-21, $10.2\pm2.5$ Gyr, and 
HAT-P-22, $12.4\pm2.6Gyr$, see Bakos et al. (2010)), CoRoT-17 has an age that is inferred 
to be $10.7\pm1.0$ Gyr. Its present bolometric luminosity is about $2.47L_\odot$, implying 
that as the star is evolving, the planet is receiving more energy. With an equilibrium 
temperature of $\sim$1600K, the planet still appears to be safe in terms of 
evaporation, however, and any possible increase of its radius should remain very 
limited.

A star with $\sim$ 1.0$M_\odot$ spends approximately 10-11 Gyr on the main sequence. 
Stellar evolution studies predict that a star similar to CoRoT-17 reaches its present 
radius at $\sim$ 90\% of its main-sequence lifetime (Hurley et al. 2000), depending on 
its metallicity and on its exact mass. The present orbital radius of the planet is 
$9.95R_\odot$, whereas we expect the star in its giant phase to reach well over 
$150R_\odot$. Even if the star looses mass owing to winds, the engulfment of the 
planet seems inevitable (see e.g. Rasio et al. 1996).

%
%____________________________________________________________________________
\begin{acknowledgements}

The team at IAC acknowledges support by grant ESP2007-65480-C02-02 of
the Spanish Ministerio de Ciencia e Innovaci{\'o}n.
This research has made use of the ExoDat database, operated at
LAM-OAMP, Marseille, France, on behalf of the CoRoT/Exoplanet
program. 
This publication makes use of data products from the Two Micron All
Sky Survey, which is a joint project of the University of
Massachusetts and the Infrared Processing and Analysis
Center/California Institute of Technology, funded by the National
Aeronautics and Space Administration and the National Science
Foundation.
This research has made use of NASA's Astrophysics Data System.
TRAPPIST is a project funded by the Belgian Fund for Scientific Research (Fond 
National de la Recherche Scientifique, FNRS) under the grant FRFC 
2.5.594.09.F, with the participation of the Swiss National Science 
Fundation (SNF).
M. Gillon and E. Jehin are FNRS Research Associates. M. Endl, W.D. 
Cochran  
and P.J. MacQueen were supported by NASA Origins of Solar Systems grant 
NNX09AB30G. 
The German CoRoT Team (TLS and the University of Cologne) acknowledges DLR 
grants 50 OW 204, 50 OW 0603 and 50QP07011. 

%http://www.eso.org/sci/observing/policies/publications.html
%Please notify Uta Grothkopf at esodata@eso.org upon acceptance or
%publication of a paper based on ESO data, including the bibliographic
%reference (article title, authors, journal title, volume, year,
%pages) and the program IDs of the data used. 

\end{acknowledgements}

%
%________________________________________________________________
%\bibliographystyle{bibtex/aa}

%\bibliography{bibl}

\end{document}